\documentclass[12pt]{article}
\usepackage{amssymb}
\newcommand{\vek}[1]{\mbox{\bf #1}}
\begin{document}
\hspace*{\fill} LMU--TPW--98/06\\
\hspace*{\fill} hep-ph/9803209 \\[3ex]

\begin{center}
\Large\bf
Confinement from a massive scalar in QCD 
\end{center}
\vspace{2ex}
%\section*{}
\normalsize \rm
\begin{center}
{\bf Rainer Dick}\\[0.5ex] {\small\it
Sektion Physik der Ludwig--Maximilians--Universit\"at M\"unchen\\
Theresienstr.\ 37, 80333 M\"unchen,
Germany}
\end{center}

\vspace{5ex}
\noindent
{\bf Abstract}: 
A model is introduced with a massive scalar coupling to the Yang--Mills term
in four--dimensional gauge theory.
It is shown that the resulting potential of colour sources consists
of a short range Coulomb interaction and a long range confining part. Far
away from the source the scalar vanishes $\sim r^{-1}$ 
while the potential
diverges linearly $\sim r$.
Up to an $N_c$--dependent factor of order 1 the tension parameter in the
model is $gmf$, where $m$ denotes the mass of the scalar and $f$ is
a coupling scale entering the scalar--gluon coupling.

\newpage  
\noindent
{\bf 1.\ }Recently it was observed that a string inspired
coupling of a massless dilaton to gauge fields yields a linearly increasing vector
potential from pointlike colour sources, if a logarithmic divergence of the dilaton
at infinity is permitted
 \cite{rd1,rd2}. This motivated me to construct a direct coupling of a massive scalar to
chromo--electric and magnetic fields subject to the requirement that the Coulomb problem
still admits an analytic solution, but now with a mass term.
The model for the scalar--gluon coupling
that emerged from this endeavour is
\begin{equation}\label{lagscal}
{\cal L}=-\frac{1}{4}\frac{\phi^2}{f^2+\beta\phi^2}F_{\mu\nu}{}^jF^{\mu\nu}{}_j
-\frac{1}{2}\partial_\mu\phi\cdot\partial^\mu\phi-\frac{1}{2}m^2\phi^2,
\end{equation}
where $0\leq\beta\leq 1$ is a parameter and $f$ is a mass scale characterizing the strength
of the scalar--gluon coupling. 

To analyze the Coulomb problem in this theory we consider
a pointlike colour source, which 
in its rest frame is described by a current
\[
j^\mu_i=g\delta(\vek{r})C_i\eta^\mu{}_0.
\]
Here $1\leq i\leq N_c^2-1$ is an su($N_c$) Lie algebra index 
and $C_i=\zeta^+\cdot X_i\cdot\zeta$ is the expectation value of the su($N_c$) generator
 $X_i$ for a normalized spinor $\zeta$ in colour space. These expectation values satisfy
\[
\sum_{i=1}^{N_c^2-1} C_i^2=\frac{N_c-1}{2N_c},
\]
and
the equations of motion for the scalar and gluons emerging from this source
follow as in \cite{rd1}:
\begin{equation}\label{eqe}
\nabla\cdot\Big(\frac{\phi^2}{f^2+\beta\phi^2}\,\vek{E}_i\Big)=gC_i\delta(\vek{r}),
\end{equation}
\begin{equation}\label{eqphi}
\Delta\phi=m^2\phi-\frac{f^2\phi}{(f^2+\beta\phi^2)^2}\,\vek{E}^i\cdot\vek{E}_i,
\end{equation}
and $\nabla\times\vek{E}_i=0$ implies existence of 
chromo--electric potentials $\vek{E}_i=-\nabla\Phi_i$.

Clearly, the solution to the Coulomb problem (\ref{eqe},\ref{eqphi}) 
includes the case of inertial motion of the colour source through a mere Lorentz boost.

 Eq.\ (\ref{eqe}) or more generally its analog for an arbitrary spherically
symmetric colour density yields for the fields outside the density
\[
\frac{\phi^2}{f^2+\beta\phi^2}\,\vek{E}_i=\frac{gC_i}{4\pi r^2}\,\vek{e}_r,
\]
and inserting this relation in (\ref{eqphi}) yields
\begin{equation}\label{eqphi2}
\Delta\phi=m^2\phi-\frac{\mu^2}{r^4\phi^3},
\end{equation}
where the abbreviation
\[
\mu=\frac{gf}{4\pi}\sqrt{\frac{N_c-1}{2N_c}}
\]
was used.

Substituting $y(r)=r\phi(r)$ in (\ref{eqphi2}) and multiplying by
 $dy/dr$ yields the first integral
\[
\Big(\frac{dy}{dr}\Big)^2=m^2y^2+\frac{\mu^2}{y^2}+2K.
\]
This can readily be solved for arbitrary
integration constant $K$, but the
boundary condition $\lim_{r\to\infty}\phi(r)=0$ 
uniquely determines $K=-m\mu$.
This yields
\[
y^2(r)=\frac{\mu}{m}+\Big(y_0^2-\frac{\mu}{m}\Big)\exp(-2mr).
\]

Therefore, the scalar field emerging from the pointlike colour source is
\begin{equation}\label{solphi}
\phi=\pm\frac{1}{r}\sqrt{\frac{\mu}{m}+\Big(y_0^2-\frac{\mu}{m}\Big)\exp(-2mr)},
\end{equation}
while the chromo--electric potentials consist of a short range Coulomb and
a long range confining part:
\begin{equation}\label{solpot}
\Phi_i=\beta\frac{gC_i}{4\pi r}-fC_i\sqrt{\frac{N_c}{2(N_c-1)}}\,
 \ln\!\Big(\exp(2mr)-1+\frac{m}{\mu}y_0^2\Big).
\end{equation}

At large distance the scalar field vanishes
 $\sim r^{-1}$, while the chromo-electric
potential yields linear confinement, if applied 
in the framework of a reduced Salpeter or no--pair equation \cite{sovw}.\\[1ex]
{\bf 2.\ }Eq.\ (\ref{solpot}) implies that an (anti-)quark with colour 
orientation $\zeta_q$, $\zeta_q^+\cdot\zeta_q=1$,
in the field of a source of colour $\zeta_s$ (a heavy quark) sees a potential
\begin{equation}\label{interpot}
V(r)=\pm\Big(|\zeta_s^+\cdot\zeta_q|^2-\frac{1}{N_c}\Big)
\Big[\beta\frac{g^2}{4\pi r}
-gf\sqrt{\frac{N_c}{2(N_c-1)}}\, \ln\!\Big(\exp(2mr)-1+\frac{m}{\mu}y_0^2\Big)\Big],
\end{equation}
with the upper sign holding for quark--quark interactions and the lower sign
applying to quark--anti-quark interactions.
The colour factor $|\zeta_s^+\cdot\zeta_q|^2-\frac{1}{N_c}$ defines a double
cone around the direction of $\zeta_s$. This double cone has an 
angle $\tan\theta_c=\sqrt{N_c-1}$ against the symmetry axis
specified by $\zeta_s$, and separates 
domains of attraction from domains of repulsion: Quarks of colour $\zeta_q$
in the double cone are repelled and anti-quarks are attracted, while
quarks with colour outside the cone are attracted and anti-quarks are repelled.

Potentials with an $1/r$ singularity at short distances
and linear behaviour at large distances have been very successfully applied
in the investigation of the quarkonium spectrum, see e.g.\ \cite{qs}
for two of the classical
references in the field.
Phenomenological tension parameters in heavy--light meson systems
are of order $\sigma\simeq (430\,\mbox{MeV})^2$ \cite{hl}, and in the present model the
tension would be determined by the mass and coupling scale of the
scalar field according to $\sigma\simeq gmf$.

 Eqs.\ (\ref{solpot},\ref{interpot}) indicate that 
direct couplings of scalar fields to Yang--Mills terms provide
an interesting paradigm for the description of confinement
in gauge theories. 
This might be realized in QCD through a fundamental
scalar, or eventually through a low energy effective scalar degree of freedom.
The possibility 
of a fundamental scalar cannot be excluded, since 
the coupling scale $f$ might be large enough to make such a scalar 
invisible to present day experiments. On the other hand, there
exist scalar resonances in the hadronic spectrum whose r\^{o}le has not been
understood yet.

On the level of a low energy effective scalar degree of freedom, one might
speculate that the Lagrangian (\ref{lagscal}) with $\beta=0$
is realized in the low energy regime
through a QCD dilaton coupling to the trace anomaly,
while at high energies the standard QCD Lagrangian would apply.
Such a picture could be motivated, if one combines the old idea
of scalar meson dominance of the trace of the energy--momentum
tensor \cite{smd} with the QCD trace anomaly, as in \cite{BMY}.
Since the trace anomaly is proportional to the gluon condensate
this could also justify the scalar gluon coupling in (\ref{lagscal})
with $\beta=0$, and eq.\ (\ref{interpot}) then tells us how the dilaton
changes the quark interaction potential.
A disadvantage with this picture concerns the disappearance of the
one gluon exchange term in the low energy regime.

In conclusion, the derivation of (\ref{interpot}) provides a challenge to
monopole condensation as a mechanism for quark confinement, and
it seems well justified to dedicate some more efforts
to the investigation of phenomenological aspects of scalar--gluon
couplings of the kind described in (\ref{lagscal}).\\[1ex]

\noindent
{\bf Acknowledgement:} I would like to thank my colleagues in Munich
for continuous encouragement and support.

\end{document}